# Anisotropic thermal transport in bulk hexagonal boron nitride

Puqing Jiang,[1] Xin Qian,[1] Ronggui Yang,[1,*] and Lucas Lindsay[2,†]

[1]*Department of Mechanical Engineering, University of Colorado, Boulder, Colorado 80309, USA*

[2]*Materials Science and Technology Division, Oak Ridge National Laboratory, Oak Ridge, Tennessee 37831, USA*

**ABSTRACT**

Hexagonal boron nitride (*h*-BN) has received great interest in recent years as a wide bandgap analog of graphene-derived systems, along with its potential in a wide range of applications, for example, as the dielectric layer for graphene devices. However, the thermal transport properties of *h*-BN, which can be critical for device reliability and functionality, are little studied both experimentally and theoretically. The primary challenge in the experimental measurements of the anisotropic thermal conductivity of *h*-BN is that typically sample size of *h*-BN single crystals is too small for conventional measurement techniques, as state-of-the-art technologies synthesize *h*-BN single crystals with lateral sizes only up to 2.5 mm and thickness up to 200 μm. Recently developed time-domain thermoreflectance (TDTR) techniques are suitable to measure the anisotropic thermal conductivity of such small samples, as it only requires a small area of 50x50 μm$^2$ for the measurements. Accurate atomistic modeling of thermal transport in bulk *h*-BN are also challenging due to the highly anisotropic layered structure. Here we conduct an integrated experimental and theoretical study on the anisotropic thermal conductivity of bulk *h*-BN single crystals over the temperature range of 100 K to 500 K, using TDTR measurements with multiple modulation frequencies and a full-scale numerical calculation of the phonon Boltzmann transport equation starting from the first principles. Our experimental and numerical results compare favorably for both the in-plane and through-plane thermal conductivities. We observe unusual temperature-dependence and phonon-isotope scattering in the through-plane thermal conductivity of *h*-BN and elucidate their origins. This work not only provides an important benchmark of the anisotropic thermal conductivity of *h*-BN, but also develops fundamental insights into the nature of phonon transport in this highly anisotropic layered material.

---

[*] Ronggui.Yang@Colorado.Edu

[†] lindsaylr@ornl.gov

*This manuscript has been authored by UT-Battelle, LLC under Contract No. DE-AC05-00OR22725 with the U.S. Department of Energy. The United States Government retains and the publisher, by accepting the article for publication, acknowledges that the United States Government retains a non-exclusive, paid-up, irrevocable, world-wide license to publish or reproduce the published form of this manuscript, or allow others to do so, for United States Government purposes. The Department of Energy will provide public access to these results of federally sponsored research in accordance with the DOE Public Access Plan (http://energy.gov/downloads/doe-public-access-plan).*

# I. INTRODUCTION

Hexagonal boron nitride (*h*-BN), a layered two-dimensional (2D) material with similar lattice structure as graphite, has been frequently used as an ultraviolet light emitter,[1,2] as a solid lubricant in harsh environments,[3] and as a thermally conductive and electrically insulating filler in composites,[4] due to its unique physical properties. More recently, there is increasing interest in using *h*-BN as a substrate for graphene electronics.[5] Many unique features of *h*-BN, including its atomically smooth surface (free of dangling bonds and charge traps), very high-frequency optical phonon modes, large electronic bandgap, and especially similar lattice constants to graphene, make it an ideal substrate for graphene electronics. Graphene devices on *h*-BN substrates are shown to have the mobility and carrier homogeneity that are almost an order of magnitude better than graphene devices on $SiO_2$ substrates.[5] Moreover, graphene/*h*-BN heterostructures exhibited many interesting phenomena[6-11] and have been proposed for a wide range of applications including field-effect transistors (FET),[12-14] quantum tunneling transistors,[15] high-frequency oscillators,[16,17] tunneling diodes,[18] LEDs,[19] and solar cells.[20] Thermal management is critical in many of these BN-related applications; it is therefore of fundamental importance to develop a detailed understanding of the thermal transport properties of *h*-BN.

The anisotropic thermal conductivity of bulk *h*-BN single crystals, however, remain little studied to date, either experimentally or theoretically. When the in-plane thermal conductivity of bulk *h*-BN is referred to in the literature,[21,22] the data of a pyrolytic compression annealed *h*-BN sample measured by Sichel *et al.*[23], with a value of 390 W m$^{-1}$ K$^{-1}$ at room temperature measured using a steady-state technique, is often cited. Sichel's data, however, is almost twice as high as the data measured by Simpson *et al.*[24] using a flash technique, also with a pyrolytic compression annealed *h*-BN sample. It is unclear whether the difference between these two studies is due to the inconsistence in sample quality or measurement methods. The only measured data of through-plane thermal conductivity of bulk *h*-BN available in the literature is from Simpson *et al.*,[24] who reported it to be ~2.5 W m$^{-1}$ K$^{-1}$ at room temperature. The primary challenge in measuring the thermal conductivity of *h*-BN single crystals is that the typical size of single crystals is not large enough for many characterization techniques,[25] including the steady-state technique used by Sichel *et al.*[23] and the flash technique by Simpson *et al.*[24]. State-of-the-art synthesis techniques produce *h*-BN single crystals with lateral sizes only up to 2.5 mm and thickness up to 200 μm.[26,27] As such, the samples measured by Sichel *et al.*[23] and Simpson *et al.*[24] 40 years ago were prepared into a centimeter size by hot pressing and annealing and thus were not single crystals in a strict sense, but with porosity. On the other hand, the recent advancement of an ultrafast laser-based time-domain thermoreflectance (TDTR) method has made it possible to measure anisotropic thermal conductivity of very small-scale samples.

Likewise, it is a surprise that theoretical modeling of the thermal conductivity of bulk *h*-BN is lacking, likely due to the challenges in accurate modeling of phonon-phonon interactions in highly anisotropic materials. Strong in-plane covalent bonding of light B and N atoms give very large acoustic velocities and high frequency phonon modes of *h*-BN along the planar direction, while weak inter-planar bonding gives small acoustic velocities and low frequency modes perpendicular to the planes (see dispersion in Fig. 2). *Ab initio* calculations based on Boltzmann transport theory[28-30] in combination with density functional theory (DFT)[31,32] have demonstrated accuracy in describing thermal conductivity of a variety of bulk materials[33-35] and 2D structures[36-40]. Yet, anisotropic layered materials, including *h*-BN, have not been studied as intensively.[41-44]

In this paper, we present an integrated study of the anisotropic thermal conductivity of bulk *h*-BN single crystals from both experiments and first-principles-based calculations over the temperature range of 100 K to 500 K. Both the in-plane and through-plane thermal conductivities of bulk *h*-BN single crystals were measured by TDTR using multiple modulation frequencies and calculated from a full numerical solution of the Peierls-Boltzmann phonon transport equation. Our measurements and calculations agree favorably with each other. We also develop fundamental insights into the nature of phonon transport in this highly anisotropic layered 2D material.

## II. METHODOLOGIES

### A. Experimental details

The *h*-BN single crystal (grade A) samples, which were purchased from SPI Supplies®, are 1-mm-sized flakes with a thickness >10 μm (see Fig. 1(b) for an optical image of a typical sample). Microscopic images of the samples demonstrate atomically flat surfaces and large domain sizes >50 μm (see Fig. 1(c)). The *h*-BN samples are of high purity, as confirmed by the sharp peak in the Raman spectrum provided by the vendor (see Fig. 1(d)). To prepare the samples for measurements, the small *h*-BN crystals were first attached to a large Si wafer using a carbon tape. The first few layers of *h*-BN were then removed by exfoliation with scotch tape immediately before the deposition of a 100 nm Al film using an e-beam evaporator. The Al film serves as a transducer for the thermal conductivity measurements.[45]

We measure the anisotropic thermal conductivities of *h*-BN using ultrafast laser-based time-domain thermoreflectance.[46] In our TDTR measurements, a train of laser pulses at an 81 MHz repetition rate is split into a pump path and a probe path. The pump pulses are modulated at a frequency in the range 1-10 MHz using an electro-optic modulator, while the probe pulses are delayed with respect to the pump via a mechanical delay stage. Both the pump and probe beams are then directed into an objective lens and focused on the sample surface. The modulated pump beam induces heating on the sample surface, with the resulting temperature change being detected by the probe beam via thermoreflectance, $\Delta R = (dR/dT)\Delta T$. A fast-response photodiode detector collects the reflected probe beam and converts the optical signals into electrical signals. A ratio-frequency lock-in amplifier picks up the thermal signal, which contains an in-phase component $V_{in}$ and an out-of-phase component $V_{out}$. The ratio of $V_{in}$ and $V_{out}$, $R = -V_{in}/V_{out}$, recorded as a function of delay time, is used to extract thermal properties of the sample through the best model fit of the measured signals.[47]

TDTR can measure both the in-plane and the through-plane thermal conductivities of the sample under different heat transfer configurations, achieved by changing the relative sizes of the laser spot $w_0$ and the in-plane thermal penetration depth $d_{p,r}$, defined as $d_{p,r} = \sqrt{K_r/\pi f C}$, where $K_r$ and $C$ are the in-plane thermal conductivity and volumetric heat capacity of the sample and $f$ is the modulation frequency. When TDTR experiments are conducted under the condition of $w_0 \gg d_{p,r}$, the heat flow is mainly one-dimensional along the through-plane direction and the TDTR signals are mainly affected by the through-plane thermal conductivity $K_z$ of the sample. When TDTR experiments are conducted under the condition of $w_0 \leq d_{p,r}$, the heat flow is three-dimensional and the TDTR signals depend on both $K_r$ and $K_z$ of the sample. Note that $w_0$ here is the root-mean-square (RMS) average of the $1/e^2$ radii of the pump and probe spots on the sample surface. Previously, we developed a variable spot size TDTR approach to measure anisotropic thermal conductivity.[46] In the variable spot size TDTR approach, different heat transfer regimes are achieved by varying the laser spot size $w_0$ and fixing the modulation frequency $f$ (and consequently $d_{p,r}$) to avoid potential artifacts from frequency-dependent $K_z$ of the sample.[46]

Several materials have been found to have frequency-dependent $K_z$, notably some layered 2D materials such as $MoS_2$[48] and black phosphorus[49], and some semiconductor alloys[50,51]. However, not all layered 2D materials are found to have frequency-dependent $K_z$, with h-BN and graphite as two such examples.[46]

In this work, a fixed laser spot size of $w_0 = 10$ μm and multiple modulation frequencies from 1 MHz to 10 MHz were used to measure the anisotropic thermal conductivity of h-BN using TDTR (specifically, the five frequencies of 1, 2.1, 3.4, 5.1, and 10 MHz were chosen). While the TDTR signals are consistently sensitive to $K_z$ of h-BN at all the modulation frequencies, the TDTR signals are highly sensitive to $K_r$ only at low modulation frequencies, as depicted by the 25% bound curves shown in Fig. 1(e). We found that thermal model predictions using a single set of parameters ($K_r$ and $K_z$ of h-BN, and Al/h-BN interfacial thermal conductance $G$) can fit the TDTR signals at multiple frequencies very well, as shown in Fig. 1(e), which indicates no modulation frequency dependence as reported previously on some other materials.[48,50,51] From this set of TDTR data, the thermal properties are extracted to be $K_r = 420$ W m$^{-1}$ K$^{-1}$, $K_z = 4.8$ W m$^{-1}$ K$^{-1}$, and $G = 60$ MW m$^{-2}$ K$^{-1}$ for single crystalline h-BN at room temperature.

Such a lack of frequency dependence is consistently evident in TDTR measurements of h-BN at other temperatures as well, which makes the data reduction much simpler. The multiple-frequency TDTR measurements of h-BN were repeated over a temperature range from 100 K to 400 K. A least squares algorithm developed by Yang et al.[52] is used to simultaneously determine $K_r$, $K_z$, and $G$ and estimate their uncertainties at each temperature. Figure 1(f) shows the confidence ranges of $K_r$, $K_z$, and $G$ of h-BN at room temperature when these parameters are simultaneously determined from the TDTR data at multiple frequencies. The uncertainties of $K_r$, $K_z$, and $G$ of h-BN at room temperature are found to be 11%, 13%, and 33%, respectively.

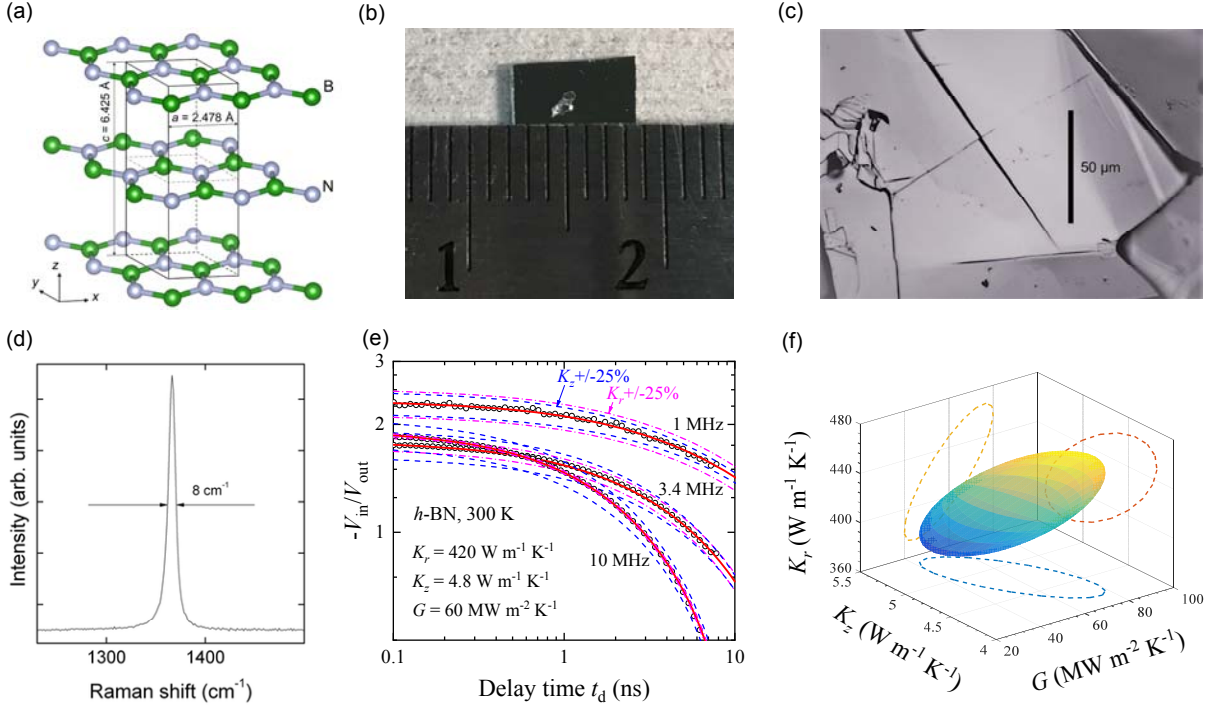

FIG. 1 (Color online) (a) Layered crystalline structure of hexagonal boron nitride (*h*-BN). The interlayer spacing is 3.213 Å and the in-plane lattice constant is equal to 2.478 Å from our first-principles calculations, which are 1% and 3.5% smaller than measurements,[53] respectively. (b, c) Optical images of an *h*-BN sample sitting on a silicon substrate. (d) Raman spectrum of *h*-BN. (e) Representative data fitting of TDTR measurements of *h*-BN at 300 K using a laser spot size $w_0 =$ 10 μm and multiple modulation frequencies between 1-10 MHz. A single set of parameters $K_r$, $K_z$ and $G$ can be used to fit the measured data at all the modulated frequencies. (The data measured at 2.1 MHz and 5.1 MHz can be fitted equally well but are not shown here for clarity.) The dashed curves and dash-dotted curves are 25% bounds on the best fitted $K_z$ and $K_r$ values, respectively, as a guide to the reading of the sensitivity of the signals. (f) Confidence ranges of $K_r$, $K_z$, and $G$ when these parameters are simultaneously determined from the multiple-frequency TDTR measurements of *h*-BN are presented in (e).

### B. Computational details

Our calculations are based on a full solution of the Peierls-Boltzmann transport equation[28-30] with harmonic (for phonon frequencies) and anharmonic (for phonon interactions) interatomic force constants (IFCs) calculated from DFT.[31,32] More specifically, we used the plane-wave based Quantum Espresso package[54,55] within the local density approximation (LDA) using Perdew-Zunger exchange correlation[56] and norm-conserving von Barth Car pseudopotentials[57] to represent the core electrons. The bulk *h*-BN system was relaxed with the AA' layer stacking configuration (see Fig. 1(a)) using 12x12x8 electronic integration grids and a 110 Ryd energy cutoff. This gives structural lattice constants $a$ = 2.478 Å and $c$ = 6.425 Å. These parameters are 1% and 3.5% smaller than measurements,[53] respectively, as LDA calculations typically overbind the atoms.[58] DFT van der Waals corrections were not included as LDA calculations of the phonon dispersions agree well with measured data (see Fig. 2). The harmonic IFCs, high-frequency dielectric tensor and Born effective charges used to construct the dynamical matrices that determine phonon frequencies were calculated from density functional perturbation theory[59] using an 8x8x6 integration mesh. Third-order anharmonic IFCs were built from numerical derivatives of the interatomic forces from Γ-point-only electronic structure calculations of perturbed 200 atom supercells with interactions restricted to unit cell atom neighbors within 2.8 Å for atoms in the same plane and within 4.2 Å for atoms in neighboring layers.

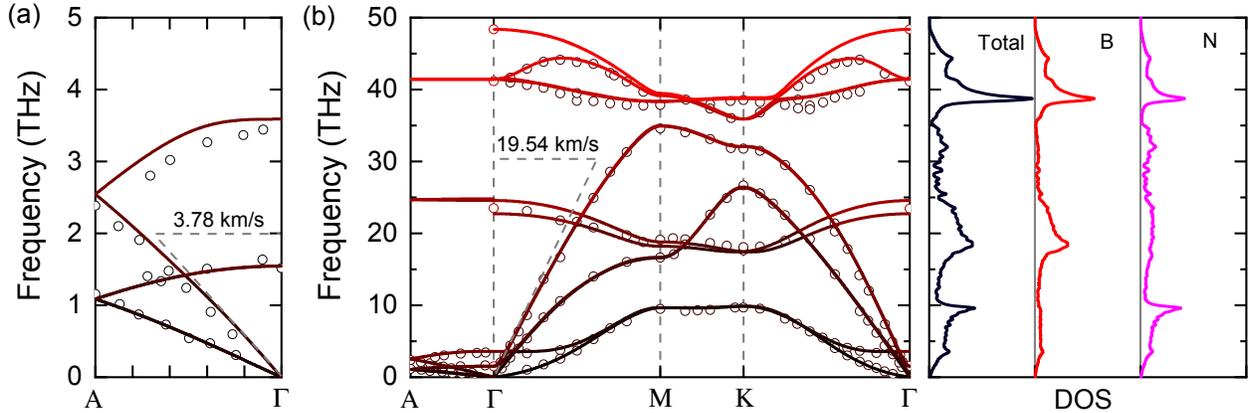

FIG. 2 (Color online) (a) and left panel of (b): Calculated phonon dispersion of *h*-BN in high-symmetry directions, which agree well with literature measurements.[60-62] Right panel of (b): Total density of states (DOS) and projected density of states on boron (B) and on nitrogen (N) sites.

## III. RESULTS AND DISCUSSION

Figure 3 summarizes the in-plane and the through-plane thermal conductivities of single crystalline $h$-BN from both TDTR measurements and first-principles predictions over the temperature range of 100-500 K, along with the literature data from Sichel et al.[23] and Simpson et al.[24] for comparison. Calculations of both isotopically pure and natural $h$-BN are presented to demonstrate the effect of phonon-isotope scattering in determining the thermal conductivity. All measurements were performed on natural $h$-BN crystal samples from SPI. The measured results and the theoretical calculations compare relatively well. At room temperature, the TDTR measurements give $K_r$ = 420±46 W m$^{-1}$ K$^{-1}$ and $K_z$ = 4.8±0.6 W m$^{-1}$ K$^{-1}$, while the first-principles method predicts $K_r$ = 537 W m$^{-1}$ K$^{-1}$ and $K_z$ = 4.1 W m$^{-1}$ K$^{-1}$ for natural $h$-BN. Our measured $K_r$ of $h$-BN compare favorably with the measurements by Sichel et al.[23] over the entire temperature range; however, both $K_r$ and $K_z$ are nearly twice larger than those measured by Simpson et al.[24]. Decreasing in temperature from 400K, both measurements and calculations demonstrate typical increasing $K_r$ as intrinsic phonon-phonon interactions become weaker. However, the measured $K_r$ value peaks at ~200 K and begin to decrease with decreasing temperature, while calculations continue to increase with $1/T$ behavior. This discrepancy is likely due to extrinsic scattering (e.g., defects, inclusions, grain boundaries, etc.) that provide relatively stronger thermal resistance at lower temperatures as intrinsic scattering becomes weak.

An interesting phenomenon is revealed by the temperature dependence of $K_z$ of $h$-BN in Fig. 3. While $K_r$ of $h$-BN exhibits a $1/T$ dependence as expected in the high-temperature range > 300 K,[63] the $K_z$ temperature dependence is relatively flat, not following the usual trend, in both our measurements and calculations. Furthermore, calculated results in Fig. 3 show a significant reduction in $K_z$ due to phonon-isotope scattering at higher temperatures > 300 K (increased spacing between the pure and natural $h$-BN curves) than observed at lower temperatures. The opposite behavior (stronger phonon-isotope scattering at lower temperatures) is expected and seen in the temperature dependence of $K_r$. These unusual behaviors of $K_z$ of $h$-BN can be understood in terms of the relatively important high-frequency phonons in determining the overall through-plane conductivity of $h$-BN, as discussed in detail below.

Figure 4 gives the contributions of conductivity from the six highest frequency branches and separately the six lowest frequency branches to $K_r$ and $K_z$ of $h$-BN as a function of temperature for both isotopically pure and natural $h$-BN. Note that some acoustic modes with frequencies > 25 THz (see Fig. 2(b)) give contributions to the high-frequency grouping. For both $K_r$ and $K_z$ of $h$-BN, the low-frequency contributions decrease as $1/T$ while the high-frequency contributions increase with increasing temperature. As temperature increases, more high-frequency phonons are thermally excited, resulting in enhanced scattering of the low-frequency modes and with relatively increased contributions from high frequency modes themselves as well. Note that at very high temperatures, the high-frequency conductivity contributions also decay as $1/T$ (see the peak behavior in high-frequency contributions in Fig. 4(a)). Across the entire temperature range of 100-1000 K, $K_r$ is dominated by the contributions from the low-frequency modes. On the other hand, due to the weak interlayer coupling in $h$-BN, the low-frequency modes have small velocities and thus their contributions to $K_z$ become comparable to or even lower than those of the high-frequency modes at higher temperatures. As the high-frequency modes become more important in $K_z$ at high temperatures, so does the phonon-isotope scattering that these modes are more susceptible to (considering the fourth power frequency dependence of the phonon-isotope scattering rate $1/\tau_{ph-iso} \sim \omega^4$). Therefore, the unusual temperature and isotope behaviors of $K_z$ of

*h*-BN at high temperatures >300 K are due to the relatively small and decreasing conductivity contributions from the low-frequency modes and the increasing contributions of the high-frequency modes as temperature increases.

For a wide variety of applications of *h*-BN in nano-devices, feature size relative to the heat-carrier mean-free-paths (MFPs) is of practical importance as it affects the effective thermal conductivity of nano-sized *h*-BN. Figure 5 gives the normalized cumulative thermal conductivity of *h*-BN for both the in-plane and through-plane directions as a function of phonon MFP at different temperatures: 100 K, 200 K and 300 K. At room temperature (300K), phonons with MFPs in the range 3-90 nm contribute ~80% of the total through-plane thermal conductivity (this corresponds to the accumulation curve between the horizontal dashed lines in Fig. 5). These MFPs in the through-plane direction are an order of magnitude smaller than the MFPs of phonons in the in-plane direction. Based on this MFP spectrum for *h*-BN, we expect that the room temperature $K_z$ of *h*-BN films depend on the film thickness for systems with thickness <100 nm and the $K_r$ of *h*-BN crystals should be reduced from the intrinsic value with grain sizes <1 µm.

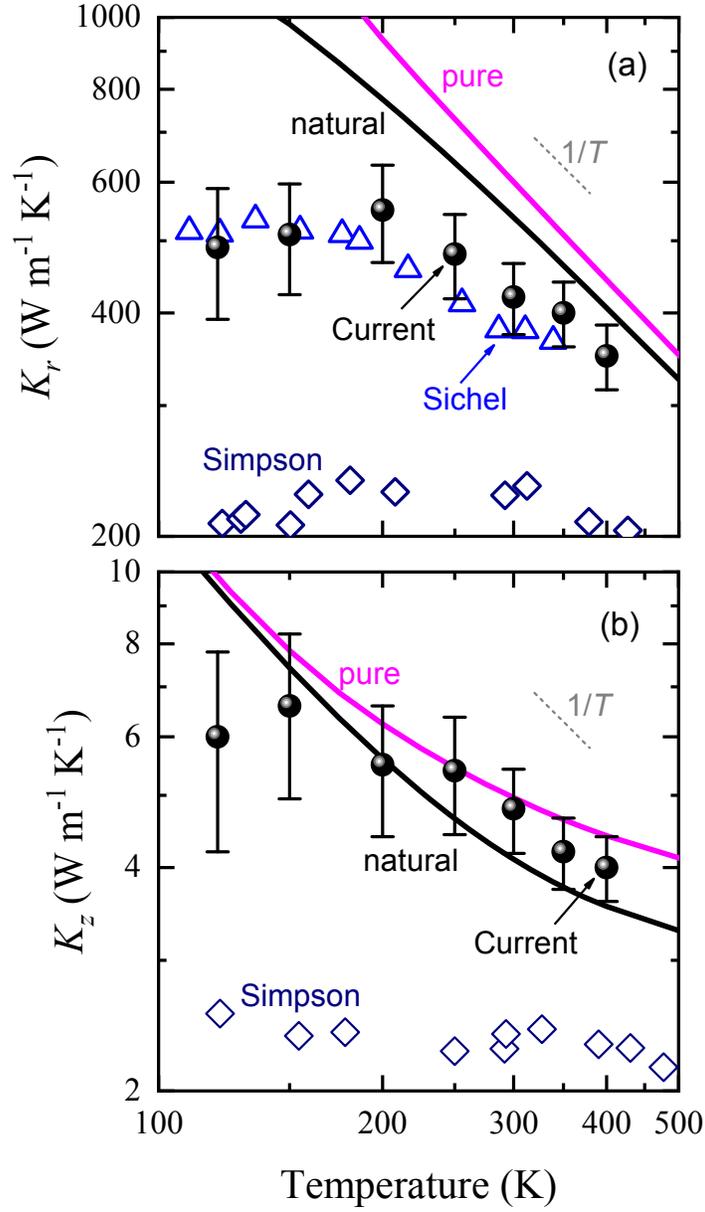

FIG. 3 (Color online) (a) In-plane and (b) through-plane thermal conductivities of $h$-BN from both our measurements (solid symbols) and first-principles calculations (solid curves) as a function of temperature, compared with measurements in the literature by Sichel *et al.*[23] and Simpson *et al.*[24]

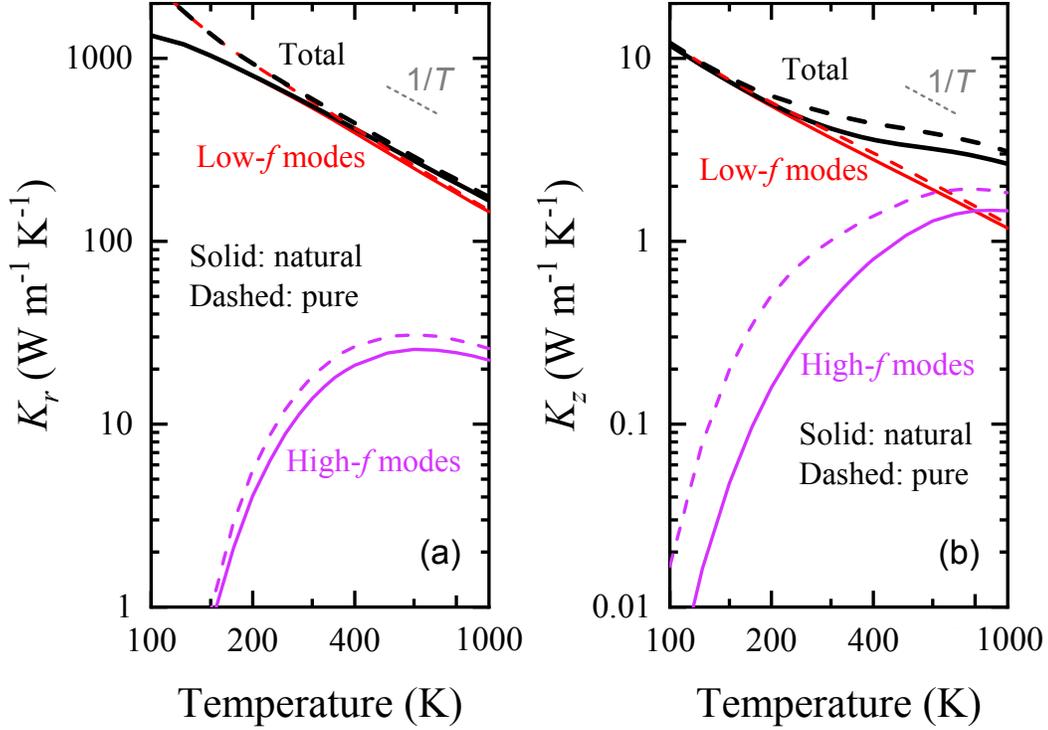

FIG. 4 (Color online) Calculated contributions to $K_r$ and $K_z$ from the six lowest frequency branches (red curves) and the six highest frequency branches (pink curves) (see Fig. 2(b) for phonon branches) as a function of temperature. Dashed curves give conductivity contributions to isotopically pure $h$-BN, while solid curves give contributions to $h$-BN with naturally occurring isotope concentrations.

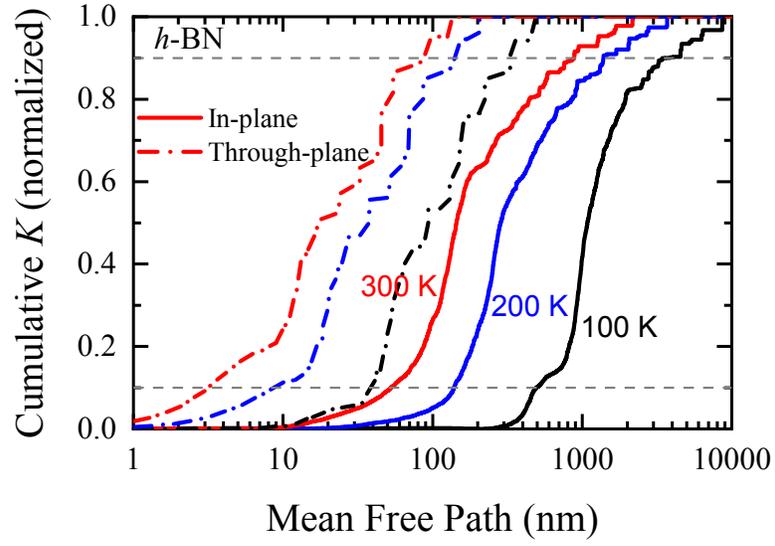

FIG. 5. (Color online) Normalized calculated cumulative thermal conductivity of *h*-BN as a function of phonon mean free path at 100 K, 200 K, and 300 K.

## IV. CONCLUSION

In summary, this work presents an integrated study of the anisotropic thermal conductivity of bulk *h*-BN single crystals using both the time-domain thermoreflectance techniques and the first-principles calculations over the temperature range of 100 K to 500 K. The TDTR measurements were conducted with multiple modulation frequencies, and the calculations were from a full numerical solution of the Peierls-Boltzmann phonon transport equation with interatomic forces from density functional theory calculations as input. Our measured and calculated temperature-dependent anisotropic thermal conductivities agree favorably well. We observed unusual behaviors in the temperature-dependent $K_z$ of *h*-BN at high temperatures >300 K, which are attributed to the relatively low contributions from low-frequency modes in the through-plane thermal conductivity of *h*-BN at high temperatures. This work not only provides an important benchmark of the anisotropic thermal conductivity of *h*-BN but also develops fundamental insights into the nature of phonon transport in this highly anisotropic layered material.

## ACKNOWLEDGEMENTS

P.J., X.Q., and R.Y. acknowledge support from NSF Grant No. 1511195 and DOE Grant No. DE-AR0000743. L. L. acknowledges support from the U. S. Department of Energy, Office of Science, Office of Basic Energy Sciences, Materials Sciences and Engineering Division.